\documentclass[journal]{IEEEtran}
\usepackage[dvips]{graphicx}
\usepackage[noadjust]{cite}

\begin{document}

\title{Cloud Networking Formation in CogMesh Environment}

\author{Tao~Chen,~\IEEEmembership{Member,~IEEE,} Honggang~Zhang,~\IEEEmembership{Member,~IEEE,} Marcos~D.~Katz,~\IEEEmembership{Member,~IEEE}
\thanks{T. Chen and M. Katz are with VTT, Finland.}
\thanks{H. Zhang is with Zhejiang University, China.}}

\maketitle

\begin{abstract}
As radio spectrum usage paradigm moving from the traditional
command and control allocation scheme to the open spectrum
allocation scheme, wireless networks meet new opportunities and
challenges. In this article we introduce the concept of cognitive wireless mesh (CogMesh) networks and address the unique problem in such a network. CogMesh is
a self-organized distributed network architecture combining cognitive technologies with the mesh structure in order to provide a uniform service platform over a wide range of networks. It is based on dynamic spectrum access (DSA) and featured by self-organization, self-configuration and self-healing. The unique problem in CogMesh is the common control channel problem, which is caused by the opportunistic spectrum sharing nature of secondary users (SU) in the network. More precisely, since the channels of SUs are fluctuating according to the radio environment, it is difficult to find always available global common control channels. This puts a significant challenge on the network design. We develop the control cloud based control channel selection and cluster based network formation techniques to tackle this problem. Moreover, we show in this article that the swarm intelligence is a good candidate to deal with the control channel problem in CogMesh. Since the study of cognitive wireless networks (CWN) is still in its early phase, the ideas provided in this article act as a catalyst to inspire new solutions in this field.
\end{abstract}

\section{Introduction}
Radio spectrum usage is undergoing a paradigm shift from the
traditional command and control allocation to the DSA \cite{report2002dn}. \textit{Cognitive radio} (CR) is a promising approach to achieve open spectrum sharing flexibly and efficiently \cite{haykin2005crb}. The research on CR has already penetrated into different types of wireless networks, and covered almost every aspect in wireless communications \cite{akyildiz2006ngd}. Following the CR, the concept of CWN comes out with the emphasis on the network-wide cognition and adaptation \cite{thomas2006cna}-\cite{bourse2004ere}.

Although a CWN may not rely on CR technologies solely, it is a common assumption that CWNs are CR based to some extent and use DSA as the spectrum access scheme. We follow this assumption in the article. In a DSA based network the SUs of the spectrum opportunistically access the spectrum based on the activities of the \emph{primary users} (PU) as well as the radio environment. The definitions of PU and SU can be found in \cite{akyildiz2006ngd}. For the deep understanding of DSA, please refer to the survey of \cite{zhao2007dsa}.

In this article, starting from our previous related work \cite{chen2007cognet},\cite{chen2008coconet}, we first introduce the concept of \emph{CogMesh}, which is defined as a self-organized distributed network architecture combining cognitive technologies with the mesh structure in order to provide a uniform service platform over a wide range of networks. A network based on this architecture is featured by self-organization, self-configuration and self-healing. It shows similarities with wireless ad hoc networks on the aspects of distributivity and self-organization. However, a CogMesh network is more flexible on spectrum, energy and network resource usage, therefore being superior to wireless ad hoc networks on performance and resource efficiency. We call the environment where CogMesh networks are operated the CogMesh environment.

Considering that at present the study on CWNs is still on its early phase, it is interesting to show readers the unique problems in the CogMesh environment and potential solutions. We identify the control channel as one of the main challenges in CogMesh. Indeed, in the CogMesh environment, because SUs opportunistically share spectrum with PUs, the network cannot rely on a global common control channel for coordination. This is different from conventional wireless networks where the common control channels are usually assumed.

In this article, we will analyze the common control channel problem in CogMesh and propose feasible solutions. For a distributed network it is desirable that the nodes share a common control channel in order to provide reliable and efficient communications, and reduce the control overhead. We first propose a control cloud concept for the control channel selection. A \emph{control cloud} is a group of connected SUs that share a common control channel. Furthermore, one can find some similarities between CogMesh networks and the collective behavior of social insects. From such an analogy, we introduce the swarm intelligence mechanism into CogMesh and use distributed algorithms to form the control clouds as large as possible. A larger control cloud means more nodes sharing the same control channel. Then, based on the control clouds, a cluster based network formation scheme is employed to further consolidate the spectrum management. The main advantage of the swarm intelligence based control channel selection and cluster based network formation is their adaptability to the radio and network environment change, which is important for CWNs.

In the remainder of the article, we will first introduce the concept of CogMesh, and discuss its differences with conventional wireless networks. Then we will describe the control channel problem in CogMesh, and provide our solutions. A wireless MAC protocol tailored for CogMesh will be given to explain how the network discovery and cluster formation are performed. We also give the simulation results illustrating the behavior of the proposed solutions. Finally, we draw the conclusion.
\section{Concept of CogMesh}
\begin{figure}[htbp]
\begin{center}
\includegraphics[width=0.45\textwidth]{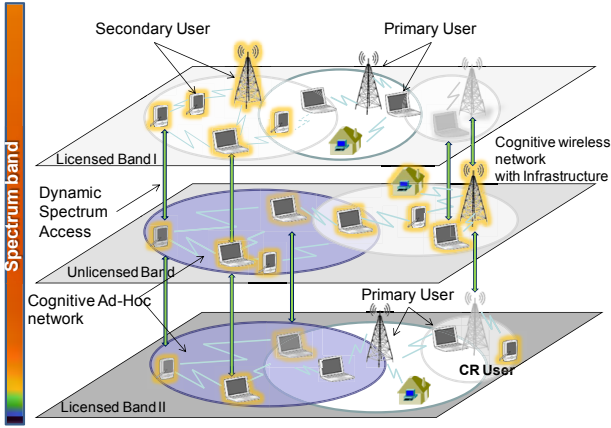}
\end{center}
\caption{A CogMesh network.} \label{fig-mesh_netw}
\end{figure}

CogMesh, as shown in Fig.~\ref{fig-mesh_netw}, is a flexible network architecture exploiting a mesh topology to integrate heterogenous wireless networks under a uniform but loosely organized control plane. It combines the advantages of CR systems and autonomous networks in a seamless way with the aim to provide a flexible network platform adaptive to a variety of existing and emerging services.

CogMesh is a wider concept than CR since the main concern of CR is the awareness, understanding and adaptation of radio resources, such as spectrum, time, space and power. Since, for instance, the power control is actually a network level problem \cite{kawadia2005pap}, without the necessary support from the network level, the flexibility brought by CR is limited. Moreover, the new open spectrum access paradigm creates a wireless ecosystem in which sub-systems work in a highly coupled way. It requires new network design principles to fully release the power of this new wireless ecosystem.

As we know, the wireless medium is unstructured in nature. The mesh structure is the natural way to match this characteristic as it provides choices to interconnect wireless devices in every possible way. It not only means the coverage extension of wireless networks, but also acts as a method to efficiently utilize resources among networks. The mesh here means the network on demand based on whatever and whenever services may require. Thus, the network topology control results into a joint optimization process as a function of service requirements, conditions of radio and network environments to ultimately fulfil the service goals and resource efficiency. In this process the cognition plays a fundamental role.

CogMesh is a rather open architecture having various forms: it can be an integration of different centralized wireless networks through a mesh structure; it can be an ad hoc network built upon the DSA paradigm; or it can be a combination of aforementioned centralized and ad hoc networks. The common features of CogMesh networks are the self-organization, self-optimization and self-healing capabilities.

\begin{figure}[htbp]
\begin{center}
\includegraphics[width=0.5\textwidth]{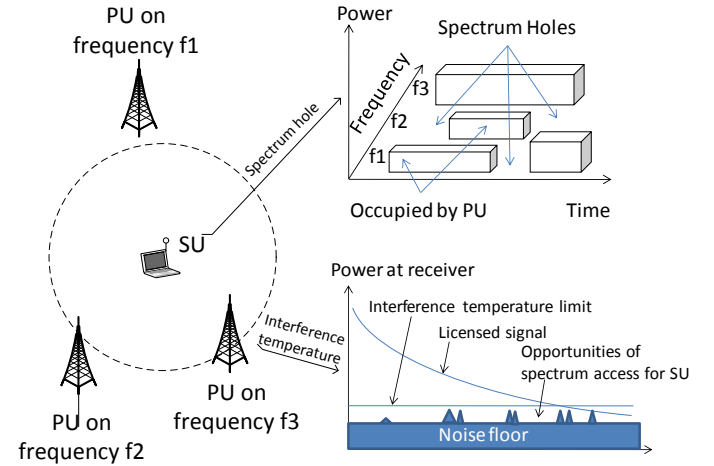}
\end{center}
\caption{Spectrum hole and interference temperature.} \label{fig-spectrum_holes}
\end{figure}

From the spectrum access perspective, a CogMesh network is typically formed by PUs and SUs of the spectrum. A SU is allowed to access a spectrum band only when causing tolerable interference to the adjacent PUs on that frequency band. We call that frequency band a \emph{spectrum hole}, which is defined as a piece of spectrum not occupied by any PU at a given time in a given geographic area. The tolerant interference of PUs can be well described by the concept of \emph{interference temperature} \cite{haykin2005crb}, which is a metric used to quantify the interference in a radio environment. Although having been temporarily abandoned by the FCC spectrum policy task force, interference temperature remains providing an accurate measure at a receiver for the acceptable level of RF interference in the spectrum band of interest, therefore playing an important role in the opportunistic spectrum sharing. The interference temperature limit of a PU serves as a cap placed on potential RF energy that could be introduced by SUs on a given band. The concepts of the spectrum hole and the interference temperature are illustrated in Fig.~\ref{fig-spectrum_holes}.

According to the interference temperature limits of PUs on different frequency bands, which are pre-specified, spectrum holes are detected through the spectrum sensing processes performed by SUs. Communications of SUs are conducted through channels extracted from those spectrum holes. Note that the spectrum hole and channel are different concepts: a spectrum hole is a continuous spectrum space of any size, while a channel is a pre-specified spectrum agreed by communication entities. According to the specification on the size of channels, a spectrum hole may hold zero to multiple channels. For each SU, the outcome of the spectrum sensing is the available channel set from multiple spectrum holes. From the channel set one channel is used as the control channel for the network coordination. Due to the complexity of the radio and network environment in CogMesh, the control channel problem becomes a prominent one. It is the focus of this article to analyze this problem and show potential solutions.

\section{Studied Scenario of CogMesh}
Under the self-organization framework of CogMesh, there are two basic forms of networking: centralized and ad hoc networking. To study the control channel problem, we focus our attention on the ad hoc networking of CogMesh since it represents the nature of self-organization and self-reconfiguration. By regarding the centralized networking as a special form of clustering, we are able to study CogMesh as a distributed network in a wide sense, especially from the control viewpoint. Therefore in this article the studied scenario of the CogMesh is the following: the SUs of CogMesh form an ad hoc network; the SUs coexist with PUs and opportunistically share the spectrum; no global common channel exists for the control purpose of SUs; by using local control channel, SUs forms a multi-hop ad hoc network. As we can see, the scenario becomes a multi-hop ad hoc network under the DSA scenario.
\subsection{Multi-channel Wireless Systems}
A CogMesh network can be modeled as a multi-channel wireless system, in which the available channels of a node vary during the life time of the node. The network coordination over multi-channel systems has been an interesting research problem. Proposals for conventional multi-channel wireless systems usually assume channels are available all the time. Two different assumptions are used when designing conventional multi-channel wireless systems: using single transceiver or using multiple transceiver. The former is the typical configuration of current wireless devices. Only one transceiver on the device means it is no possible to transmit and receive simultaneously. The latter assumes at least two transceivers at one device, therefore being capable of sending and receiving on different channels at the same time.

For the single transceiver case, three basic methods are used to coordinate the multi-channel access:
\begin{itemize}
\item \textbf{Common control channel method}, in which a common control channel is used for the signaling of all nodes. The typical case is multi-channel MAC (MMAC) \cite{so2004mcm}, which defines a default control channel where all nodes must periodically switch to and synchronize for a pre-determined window of time. The channel for data transmission is negotiated in that pre-determined window.
\item \textbf{Channel hopping method}, in which every node hops its working channel according to certain pattern and has chance to meet other nodes periodically on different channels. The hop reservation multiple access (HRMA) \cite{tang1999hrm} is one instance of this method. In HRMA, all nodes hop according to a pre-defined hopping pattern. Whenever a node has a data packet to send, it exchanges control messages with the intend receiver and both remain in the same hop pattern for the entire data transmission.
\item \textbf{Home channel for receiver method}, in which a pre-defined home channel is assigned to each node, and nodes are switched to their home channels for receiving immediately when they are idle. For instance, in \cite{chaudhuri:mpm}, every node is associated with a home channel based on node's MAC address. After data transmission, a node immediately return to its home channel for incoming packets.
\end{itemize}

For the multiple transceiver case, the multi-channel access becomes simple since additional transceivers can be used for control purpose. For instance, in \cite{nasipuri1999mcm}, nodes are assumed to have as many transceivers as the number of channels, being able to listen to all those channels simultaneously. A node having data to send simply picks up an idle channel for transmission. Obviously the cost for this approach is extremely high. Other approaches use only two transceivers, one for the control and the other for the data transmission \cite{wu2000nmc}. The control transceiver always works on the default control channel to negotiate the data channels.

As we can see, the multi-channel access solutions are all based on the assumption that all channels are always available. This is not the case in CogMesh since the availability of a channel for a SU depends on the radio environment. It means that we can not use conventional multi-channel access solutions directly. Moreover, we can not assume that every node in CogMesh has multiple transceivers. Based on this, we need to develop solutions for a general case, i.e., the single transceiver working on the half duplex mode.
\subsection{Control Channel Problem in CogMesh Networks}
It is well known that the control problem is critical in distributed networks, due to the dynamics introduced by self-coordination activities. Until now the majority of proposed spectrum control protocols designed for the DSA scenario assume the availability of a common control channel \cite{akyildiz2006ngd}. For instance, Jing et al. \cite{jing2005sce} used common spectrum coordination channel (CSCC) etiquette protocol for coexistence of IEEE 802.11b and 802.16a networks; a cognitive pilot channel (CPC) concept was proposed in \cite{cordier2006rcp} for exchanging the spectrum information among nodes.

The use of the common control channel significantly reduces the complexity of the network coordination. However, the common control channels do not always exist in CogMesh, since the SUs of CogMesh utilize spectral holes for communications. Correspondingly, the topology management of CogMesh is affected by two main factors: first, the absence of a common control channel in the network; and second, the frequent topology changes according to the presence of PUs and SUs. In the CogMesh environment, SUs use local control channels for the network coordination.

However, until now only few proposals are made under the non-common control channel assumption. For instance, Zhao et al. observed that though very limited number of global common channels exist in a network, local neighbors share numerous channels with others \cite{zhao2005dcd}. They proposed a distributed grouping scheme to solve the common control channel problem \cite{zhao2005dcd}; Bian et al. \cite{bian2007seg} used the concept of  the segment, which is a group of nodes who share common channels along a routing path, to organize control channels. In \cite{chen2007cognet}, this problem was tackled by a cluster-based approach, where the local users sharing common channels form a dynamic one-hop cluster and the spectrum is managed by cluster heads.

In the following, we will use the control cloud concept and cluster based network formation to deal with the common control channel problem, and describe how the CogMesh network is coordinated for multi-hop communications based on the proposed solutions.
\section{Network Coordination in CogMesh Networks}
\subsection{Control Cloud Concept}
\label{sec_ctrlcloud}
\begin{figure}[htbp]
\begin{center}
\includegraphics[width=0.5\textwidth]{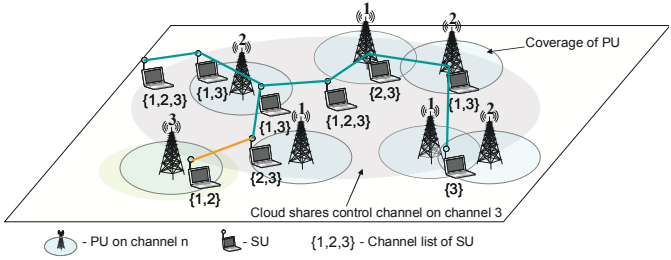}
\end{center}
\caption{Control cloud in CogMesh networks.} \label{fig-mesh_cloud}
\end{figure}
A \emph{control cloud} is a collection of neighbor SUs sharing a common control channel. We call it cloud because it may dynamically change its size according to the radio and network environment. If the whole SU network shares a common control channel, the whole network is under the control of a single control cloud. Otherwise, there are multiple control clouds separating the network.

The reason to introduce the control cloud in CogMesh is to provide a scalable control solution for CogMesh in the DSA scenario. The idea is to make control clouds grow and cover as much SUs as possible while adapting to the radio environment. The evolution of control clouds in the network results from the self-organized activities of the network. Different control clouds are interconnected through gateway nodes between the edges of the clouds.

Control clouds result from individual nodes' choices on their control channels. According to the channel quality, a node chooses a channel as its control channel, namely the \emph{master channel}, for signaling. In case that two neighbors choose different master channels, a proper listening rule can be used to perform the neighbor discovery
orderly on other channels according to their channel qualities. Once
a neighbor is detected, the proposed algorithm is run to negotiate a
common master channel among most of the neighbors. Accordingly,
channel clouds are formed and evolved with a trend to form few and large clouds as possible. Clearly, control messages running over few control channels reduces the control overhead and delay.

To setup the master channel, we use the following neighbor discovery process. Supported by the layer two or three, a SU periodically broadcasts HELLO messages over its master channel. The HELLO message includes the information of the node's master channel and all other available channels with the quantized quality values. The neighbors of the SU listen to their master channels in most of time for HELLO messages, and shift the listening to other channels with probabilities proportional to their channel qualities for a given period in a repeating manner. Once the channel information is exchanged among the neighbors, a common master channel shared by the neighbors will be negotiated by the proposed algorithm.

Therefore the control channel selection algorithm determines the formation and evolution of control clouds. We notice that there is a collective behavior in the control cloud formation, where each node makes its own decision. This effect is similar to the collective phenomenon widely seen in the biologic world. This motivates us to introduce a swarm intelligence algorithm into CogMesh targeting for control channel problems. The \emph{swarm intelligence} is a well established science biologically inspired by the collective behavior of social insects, for instance, ants or bees solving complex tasks like building nests or
foraging \cite{bonabeau1999sin}. It is based on the principle of the division of labor where the higher efficiency is achieved by specialized workers performing specialized tasks in parallel. The advantages of swarm intelligence techniques are scalability, fault tolerance, parallelism and autonomy. Swarm intelligence algorithms have been successfully employed in telecommunication networks for the performance improvement of routing protocols \cite{schoonderwoerd1997abl}, \cite{dicaro1998ads}. Recently, its applications have been found on spectrum sensing and resource allocation in CWNs \cite{Renkl2007swarm}.

In a typical swarm intelligence scenario, an agent deposits a small amount of pheromone on a trail and the trail with higher pheromone
level becomes the choice of the working trail. This distributed
optimization approach relies on the cooperation of agents to achieve
the common optimization goal with a collective complexity out of
individual simplicity. Considering each SU in a CogMesh network as a simple agent and its choice on the control channel as the pheromone, the swarm intelligence matches well the dynamics in the CogMesh network. We will describe the detail of the swarm intelligence algorithm in Section~\ref{sec_ch_sel}.
\subsection{Cluster Based Networking}
In addition to the control cloud concept, we use the cluster based networking for the network formation in CogMesh networks. The purpose of using a cluster based approach is to make the spectrum access more manageable in the DSA scenario. Managing the spectrum as a whole in the cluster reduces the control overhead when compared with the way to do it in a fully ad hoc manner, especially when SUs are coexisting with PUs. Moreover, the cluster based approach has advantages for routing in the multi-hop networking environment.

A large number of cluster formation algorithms have been proposed for ad hoc networks so far \cite{lin1997acm},\cite{amis2000mmc}. They are different on the criteria to select cluster heads. However, there are some critical problems to utilize those approaches in CogMesh networks:
\begin{enumerate}
  \item They are usually designed for the single channel case while CogMesh is a multi-channel case;
  \item They are designed for fixed network topology, and lack the capability to adapt to dynamic physical topology changes;
  \item Most of them only guarantee the network connectivity, and  therefore the result may not be optimized;
  \item In some solutions the full knowledge of the network is required, which is not realistic in CogMesh.
\end{enumerate}
As a conclusion, a different approach is needed in CogMesh.

In CogMesh, a \emph{cluster} is a group of neighbor SUs controlled by a cluster head, which is selected from that group of SUs. Normally the members of the cluster is one-hop away from the cluster head. Under the control cloud concept, the SUs in a cluster are the members of the same control cloud. Following the same rule, we call the control channel of a cluster the master channel of that cluster. The node forming the cluster becomes the cluster head, which is responsible for intra-cluster channel access control and inter-cluster communications.

\begin{figure}[htbp]
\begin{center}
\includegraphics[width=0.5\textwidth]{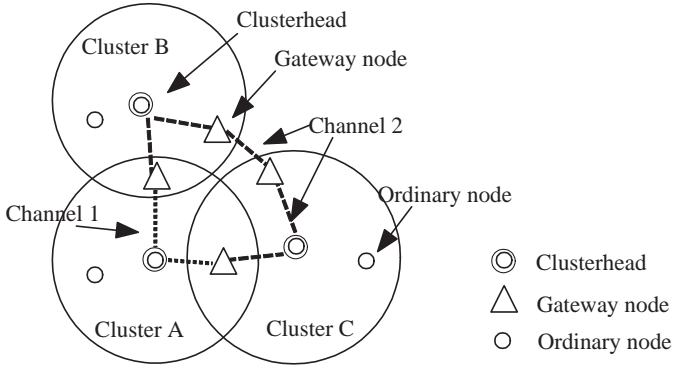}
\end{center}
\caption{Clusters interconnected by gateway nodes.}
\label{fig-bridgenode}
\end{figure}

By negotiating gateway nodes between clusters, clusters are interconnected to a large network, where multi-hop links are used to deliver data messages. A gateway node is a member of one cluster that can reach the member of another cluster. The cluster interconnection is illustrated in Fig.~\ref{fig-bridgenode}, from which, we can see that clusters are interconnected in two cases: two cluster heads are connected by one gateway node, or connected by two gateway nodes when no node is one-hop neighbor of two cluster heads. Therefore there are three types of members in a cluster: the cluster head, ordinary node, and gateway node.

In this article, we will show a specific cluster formation algorithm designed for CogMesh, with the ability to adapt to the radio and network environment. Before proceeding to the algorithm, we will introduce the MAC functions that assist the network formation in CogMesh.

\subsection{MAC Functions for Network Formation}
\begin{figure}[htbp]
\begin{center}
\includegraphics[width=0.45\textwidth]{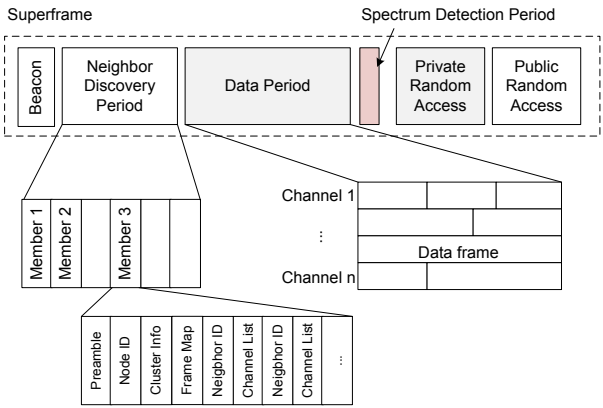}
\end{center}
\caption{Superframe structure.} \label{fig-superframe}
\end{figure}
The cluster formation and inter-cluster connection are performed distributively based on the neighbor information of nodes. We provide mechanisms in the MAC protocol to enable nodes exchange their one-hop and two-hop neighbors information, which includes neighbors's identity and their channel list. In CogMesh, a node may only know partial of its neighbors at the initial stage. The clusters are formed based on the
partial neighbor information. As nodes gradually collect more neighbor information based on the neighbor discovery mechanism, clusters are reconstructed and interconnected to a more reliable network structure.

For each cluster, channel access time is divided into a sequence of
superframes. Each superframe is divided into several periods as shown
in Fig.~\ref{fig-superframe}. The beacon period is issued by the
cluster head. It contains the time synchronization, control and
resource allocation information of the cluster. The following period
is the neighbor discovery period. It is divided into a
number of fixed length mini-slots. Each member of a cluster occupies
one mini-slot and uses it to broadcast the HELLO message, which includes its identity and one-hop neighbor list. An entry in the neighbor list includes the identity of the neighbor and its channel list. The master channel of a neighbor is indicated in the cluster list, through which a node knows how to reach the neighbor cluster. A preamble is attached at the beginning of each mini slot for other nodes identifying the
broadcasting message if they miss the beacon. Moreover, the time and
duration of random access periods in this superframe is broadcast in
the Frame Map period of its mini-slot. A neighbor of this node, once
receiving its HELLO message, has the chance to exchange its neighbor information with the node in the following random access period. The location of a member's mini-slot is announced by the cluster head in the beacon period. The number of mini-slots in a superframe is limited by a system parameter in order to avoid too many nodes crowding in one cluster.

The next period is the data period. Parallel transmissions are permitted in this period if the transmission sessions use different channels. Following the data period, an intra-cluster random access period is used for cluster members exchanging control messages. The superframe is ended by a public random access period. Its length is determined by the cluster head and announced in the beacon. This period has multiple purposes. It uses for a node joining the cluster, nodes exchanging neighbor information, or clusters exchanging control information.

Besides five main periods, there are one or several spectrum
detection periods scheduled in a superframe. During these periods,
all members of a cluster keep silence and detect spectrum holes. It
is desirable to synchronize the spectrum detection periods of
adjacent clusters so as to reduce the false alarm of the PU detection.
A false alarm occurs when a SU incorrectly
reports the presence of PUs due to the interference from
other sources. Since the superframes of different clusters are not
required to be synchronized, the location of the spectrum detection
periods varies from cluster to cluster. Even in a cluster, their
location varies from superframe to superframe.
\subsection{Neighbor Discovery and Cluster Formation}
\label{sec:FormCluster} The neighbor discovery and cluster formation
processes are introduced together since they are highly coupled. For
convenience, we say the \emph{neighbor cluster}
of a node is the cluster that the node does not belong to, but has
one-hop neighbors as its members, and the total neighbor clusters of all members of a cluster are called the cluster's neighbor clusters.

The neighbor discovery is performed during clusters' neighbor discovery periods. When a node wants to join the network, it first detects the available channels. Then it scans one of its channels for a given period of time, waiting for beacons on that channel. The node starts the scanning process from the channel with the lowest frequency, which is called the lowest channel. The scanning time on a channel is chosen
so that it exceeds the period of the longest superframe. We call a
scanning period as a scanning interval, and the first scanning
interval a new node starts as the first scanning interval. If there
is a neighbor cluster on the frequency band a node listens on, it is
able to capture its beacon during a scanning interval.

We divide the first scanning interval into three cases: no message
arrives; a beacon arrives; or neighbor messages arrive but no beacon
arrives. In the first case, the node forms a cluster on the scanning
channel and becomes the cluster head. In the second case, the node
requests to join the cluster through the public random access period of the cluster. If the cluster head accepts the request, it assigns a mini-slot to the requesting node. Starting from next superframe, the new joining node broadcasts its HELLO messages in that mini-slot. However, if there is no empty mini-slot in a cluster, the cluster head will reject the request. The requesting node then chooses the second
lowest channel to start a new scanning process, or form its own
cluster if finding the detected clusters are all full after
iterating all channels. The third case means the node has neighbor
clusters but it is two-hop away from cluster heads. The node then
records the neighbor information, and tries to exchange neighbor
information with that neighbor through the public random access period of the corresponding neighbor cluster. After that, it continues its
scanning process on the next available channel. If the node can not
find a channel satisfying the case one and two after iterating all
channels, it starts its own cluster on a randomly chosen channel.

After a node joins a cluster, it periodically chooses from its
channel list a non-master channel to scan so as to discover other
neighbor nodes. An algorithm can be developed to intelligently
choose a non-master channel according to the neighbor information
the node detects. For instance, if it discovers new two-hop neighbors
on a non-master channel, it listens on that channel first.

\begin{figure}[htbp]
\begin{center}
\includegraphics[width=0.5\textwidth]{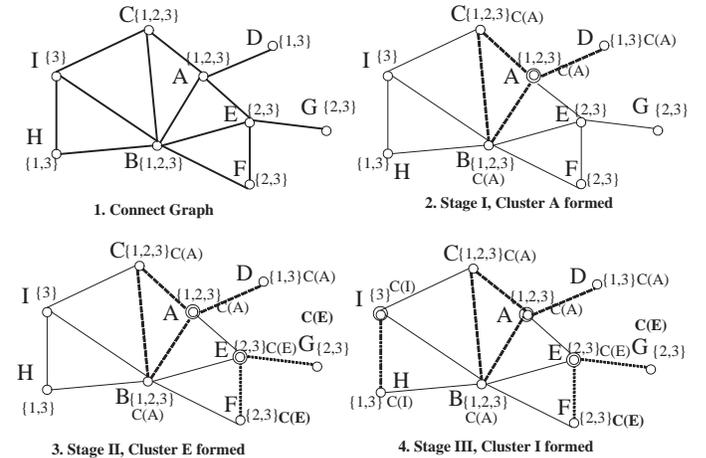}
\end{center}
\caption{Example of a cluster formation process.} \label{fig-clusterformation}
\end{figure}

Let us explain the neighbor discovery and cluster formation by an
example, as illustrated in Fig.~\ref{fig-clusterformation}. The numbers in brackets close to each node represent available channels of that
node. The smaller number represents the lower spectrum hole. We
assume the spectrum holes do not change during the cluster formation
procedure. The edge between two nodes indicates they can hear each
other. Assume the node A is the first node forming the cluster on
the channel 1. The cluster is labeled as the cluster A. Its one-hop
neighbors B, C, D listen on their lowest frequency band, i.e., the
channel 1, detect the beacon issued by the cluster A. They join the
cluster A through a corresponding association process. From the
neighbor discovery process, the node B knows the node C is its one-hop
neighbor, and the node D is its two-hop neighbor. Next, the node E, F,
G form a cluster on the channel 2. Assume the node E forms the
cluster, labeled as the cluster E. The node F, G join the cluster E
right after. The node B listens on the non-master channel 2. It
discovers E, F as its one-hop neighbors, and G as its two-hop neighbor. The cluster A and E therefore are interconnected by the node B.
Then, assume the node I forms the cluster I on the channel 3. The
node H receives B's HELLO message and detects B as its one-hop
neighbor. However, H can not receive beacons from the cluster A. It
starts a new listen process on the channel 3 and finally joins the
cluster I. The node H informs B that its new neighbor list through
the public random access period of the cluster A. The node B knows from H that there is a cluster on the channel 3. It knows the neighbors H, I on the channel 3 through a scanning process on that channel. Furthermore, the node C will know B has new neighbors H and I from the neighbor discovery period of the cluster A and finally know its neighbor I on the channel 3. At this stage, three clusters are formed, and the cluster heads has enough neighbor information for inter-cluster connection. The clusters then negotiate with each other to form a network through their public random access period.
\section{Control Cloud Formation by Swarm Intelligence Approach}
\label{sec_ch_sel} The basic idea of the swarm intelligence approach is to let a node select a channel with sufficient quality, meanwhile being preferred by most of its neighbors, as the master channel. Sufficient quality means the quality of the chosen channel ranks a higher position among all available channels. The reason to choose a better channel as the control channel is straightforward: transmission failures are reduced. The channel quality is measured in the spectrum sensing process and presented by a single value, $Q$, which is a non-negative real value inversely proportional to the accumulated interference imposed by the surrounding in a given
time window. To make it simple, we quantize the $Q$ value into several stages. The $Q$ value that falls into one stage takes a fixed value which represents that stage.

The preference of the neighbors on the master channel is
reflected by the number of neighbors who choose the same master
channel. We use an approach that takes into account the freshness of the neighbors' choice. In each node, we maintain a weight, named $W$, for each channel, updating its once receiving a HELLO message. The channel with the highest weight is selected as the master channel. In a node, the overall weights of all channel is equal to one. Periodically, a node selects its master channel according to its $W$ list.

The problem now becomes how to determine the $W$ of each
channel according to the $Q$ list of a node and its neighbors. The $W$ list, which is updated frequently according to the fluctuation of the channel quality and the choices of the neighbors on the master channels, becomes the key to reflect the radio environment and determine the master channel. We apply the swarm intelligence algorithm to update the weights.

In our network, each node acts as an agent, using the HELLO message as the pheromone to influence its neighbor nodes. A node receiving a HELLO message updates its $W$ list as follows. The channel equal to the master channel of the broadcasting neighbor receives a positive reward with an amount proportional to the difference of the master channel quality between the neighbor and the local node. All other channels receive negative rewards to make the sum of all weights remain one. This process can be mathematically presented as follows.
The parameter $W_j$, which is the $W$ value of
the channel $i$ on the node $A$, is updated by:
\begin{equation}
\label{eq_p_value_chm}W_i = W_i + r(1-W_i);
\end{equation}
where $r$ is a parameter determined by $\Delta Q$, which is the difference of the master channel quality between the neighbor sending the HELLO message and the node $A$, i.e.,
\begin{equation}
\label{eq_r} r= f(\Delta Q); \;\;\;\mbox{where } r \in [0,1]
\end{equation}
The $r$ function in (\ref{eq_r}) is a monotonically increased function. For all channels other than the channel $i$, their $W$ values on the node $A$ are updated by:
\begin{equation}
\label{eq_p_value_nchm} W_j = W_j(1-r);\;\;\; \mbox{for } W_j \in
\{W_j| j=1,...,N; \; j \ne i\}
\end{equation} where $N$ is the total number of channels on the node $A$.

This is a process in which a node persuades its neighbors to move to its master channel. The shift of a master channel happens when sufficient pheromone is accumulated on a non-master channel. On the other hand, the channel quality will be affected by PUs and thus changed over time. A node updates its channel weight list periodically according to the refreshed channel quality list. It acts as the disturbed factor to push the master channel back to the best quality channel. The amplification of disturbed factor makes the master channel evolve with the radio environment.

As we may notice, the channel weight is updated according to the difference between channel qualities. By choosing different map functions between the channel quality and the weight, we are able to control the behavior of the swarm intelligence algorithm. For instance, if we reward the channel weight proportional to the difference of channel qualities, the control cloud will keep stable under the small variation of the radio environment; if we reward the channel weight in an opposite way, the control cloud will be more sensitive to the radio environment. The example of the $r$ function can be:
\begin{equation}
\label{eq_r_value} r=[\arctan(A*\Delta Q) +B)]/C;
\end{equation}
where $A$, $B$ and $C$ are the constants affecting the
converging rate of the algorithm. (\ref{eq_r_value}) gives a smaller $\Delta Q$ more reward than a bigger $\Delta Q$. An online learning strategy can be applied here to tune the map functions so that a node can reflect its desire on either the exploitation of the most common channels or the exploration of the best quality channels.

The proposed algorithm has several advantages. First, it is independent of a specific physical and MAC layer. As a result, it can be easily integrated into heterogeneous wireless networks. Secondly, the algorithm is flexible in the sense that the parameters of the algorithm can be tuned to suit different network scenarios, for instance, adapting the HELLO message broadcasting rate to the radio environment.

Since the weight value reflects the channel quality and willingness of the nodes to utilize the channels, it has added values for clustering, routing and data transmission. The weight value for the master channel can ease the cluster management in such a network, and then improve the spectrum efficiency. A routing protocol integrating that weight value will be more intelligent to adapt to the radio environment, therefore being more flexible and robust. In addition to using the weight value to choose the control channel, the neighbor SUs can use it to select the transmission channel as well, therefore increasing the spectrum efficiency.

\section{MDS Based Cluster Reformation Algorithm}
\begin{figure*}
\begin{center}
\includegraphics[width=1\textwidth]{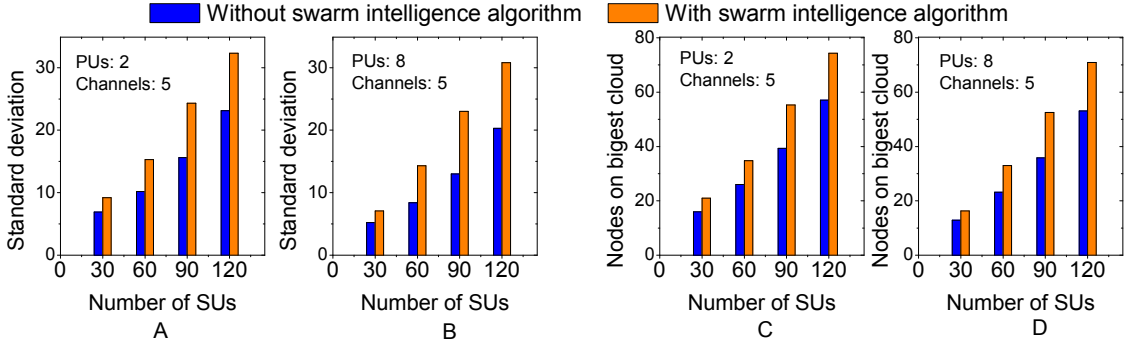}
\end{center}
\caption{Performance of proposed algorithms.} \label{fig-perf_all}
\end{figure*}
\label{sec_mds_cluster}
The cluster optimization problem can be considered as a \textit{dominating set} (DS) problem in graph theory, whose objective is to find a subset of nodes called DS with the following properties: each node is either in the DS, or is adjacent to a node in the DS \cite{bao2003tma}. In our network, the DS is the
collection of cluster heads. The cluster optimization problem is to
find a minimal dominating set (MDS) of the CogMesh network according
to its physical topology. A MDS is the minimal size DS among all possible DSs in the topology. The MDS problem is proven to be a NP-hard problem even when the complete network topology is available
\cite{amis2000mmc}. However, a sub-optimum DS can be obtained
through a local minimum election of the dominators by a heuristic
algorithm. The algorithm is run periodically and distributively on
each node and only relies on the discovered neighbor information to
determine the locally optimized cluster configuration. As a result,
the collection of cluster heads is gradually converged to a sub-optimum DS.

When the physical topology changes due to the events such as new
nodes joining the network, nodes leaving the network, or radio
environment changing, the affected nodes or clusters are
reconfigured to immediately absorb the changes. The optimization
algorithm is performed thereafter to optimize the changed physical
topology. The basic rule for the reconfiguration is when an affected
node currently belongs to no cluster, it takes action to associate
with one cluster or start a new cluster. In other cases, the
cluster heads coordinate the changes. Note that after
reconfiguration, gateway nodes of affected clusters may need
reconfiguration.

The algorithm works as follows. From the neighbor list, a node, namely the working node, obtains a node set, which includes all members of its one-hop neighbor clusters and its host cluster. It is the target node set to be optimized. A connection graph is created based on the target node set. The objective is to construct clusters based on a MDS of the graph so that the number of clusters in the target node set can be minimized.

The MDS is obtained by a heuristic algorithm  \cite{chvatal1979ghs}. The algorithm takes the multiple channels of a node into account. First, a cluster is formed by taking the working node as the cluster head and its control channel as the master channel. The one-hop neighbors of the node are assigned to the cluster if their control channels are as same as the master channel of the working node. The members of the formed cluster are eliminated from the target node set. The remaining nodes are processed as following. As in a Max Degree algorithm \cite{bao2003tma}, a node with max degree on its control channel is chosen to form a cluster with corresponding neighbors in order until all nodes join the network. Finally, the new cluster configuration comes out with the cluster heads list, the master channels and the members of each cluster. If the number of resultant clusters is smaller than the current one, the working node starts a negotiation process to reconfigure its surrounding clusters.

To start the negotiation process, the node sends rearrangement requests to the cluster heads it wants to reconfigure, indicating the gain that can be obtained from the rearrangement, and the reconfiguration instruction. The gain is the total number of clusters being reduced if the rearrangement is taken. A cluster head, once accepts the request, sends an acknowledge to the working node. The working node negotiates with the target cluster heads to complete the remaining configuration process only after receiving all acknowledges back. Otherwise, it cancels the process to avoid increasing the cluster number by an incomplete reconfiguration.

\section{Performance Analysis}
In this section we simply show the performance of swarm intelligence algorithm by simulation. The purpose is to give readers a rough idea on the behaviors of the algorithm without going deeply into the algorithm. Simulations are run under different network conditions.

We use the standard deviation of the SU number distributed on each channel to measure the trend of SUs towards sharing the master channels. A large standard deviation means the sizes of channel clouds are not evenly distributed, therefore more nodes being aggregated to few large clouds. Moreover we use the size of the largest cloud, i.e., the number of SUs in the largest cloud, to illustrate the impact of the swarm intelligence on the control cloud formation.
\begin{figure}[htbp]
\begin{center}
\includegraphics[width=0.45\textwidth]{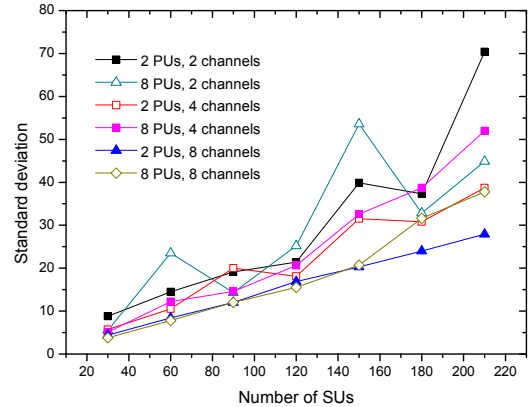}
\end{center}
\caption{Standard deviation as a function of SU
number.} \label{fig-nds_std}
\end{figure}

We first compare the performance of cluster formation with or without the swarm intelligence algorithm. The performance figures are shown in Fig.~\ref{fig-perf_all}, in which the sub-figure $A$ and $B$ show the standard deviation, and the sub-figure $C$ and $D$ show the largest control cloud size. From those figures, we conclude that the swarm intelligence algorithm forms larger control clouds, as expected.
\begin{figure}[htbp]
\begin{center}
\includegraphics[width=0.45\textwidth]{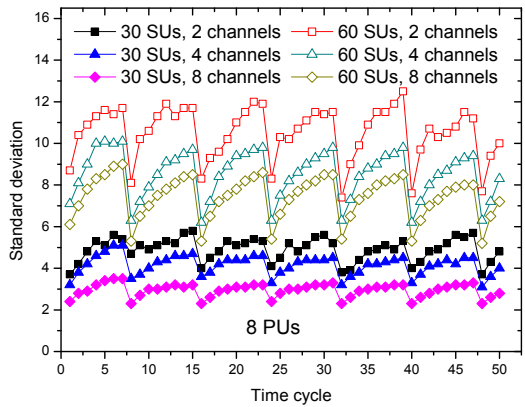}
\end{center}
\caption{Dynamic behavior of a swarm intelligence algorithm in a CogMesh network.}
\label{fig-std_dyn_8p}
\end{figure}

We then show the performance of the swarm intelligence algorithm under different SU, PU and channel settings. Fig.~\ref{fig-nds_std} shows the behavior of the algorithm corresponding to different SU populations. As seen from the figure, the standard deviation is high in all cases, meaning the algorithm works as expected. The standard deviation increases as the SUs increase. It implies more SUs are aggregated to few common master channels.

The dynamic behaviors of the algorithm are shown in Fig.~\ref{fig-std_dyn_8p}. The PUs in this simulation setup change
their operating channels periodically. As seen from this figure, in both cases after the channel fluctuation, the standard deviation turns to a high stable value shortly, meaning the SUs are aggregated to few common master channels quickly.
\section{Conclusion}
\label{sec_concl}
The DSA paradigm brings new opportunities into the wireless world. However, the benefits of the new spectrum access paradigm do not come naturally. There are plenty of challenges need to be dealt with before releasing its full power.

In this article, we introduce a new concept of the cognitive wireless network, named CogMesh, to meet those challenges. We identify that CogMesh shares similarities with conventional wireless ad hoc networks in terms of distributive control and self-organization, but at the same time there are significant difference between them since in CogMesh SUs uses opportunistic spectrum access. One of the prominent challenges in CogMesh is the control channel problem. We explain the problem in detail, and propose corresponding solutions. Inspired from the collective behaviors in the biology world, we introduce a swarm intelligence algorithm to form control clouds in CogMesh. On the top of the control clouds, we propose the cluster based network formation solution to further consolidate the spectrum management in the DSA scenario. Cluster based approach simplifies the resource management of multi-hop networking in dynamic radio environments.

The solutions provided in this article aim to provide some insights on the control problem in CWNs and show how a bio-inspired approach can be applied in CWNs. Actually the concept of CWNs opens up a new research area that covers almost every aspect of communications. Interdisciplinary research is highly beneficial in this new area. It therefore calls for the contributions from researchers with diverse background.

\bibliographystyle{IEEEtran}
\bibliography{IEEEabrv,CRNetwork}

\end{document}